\documentclass[pdflatex,sn-mathphys-num]{sn-jnl}% Math and Physical Sciences Numbered Reference Style
\usepackage{graphicx}%
\usepackage{multirow}%
\usepackage{amsmath,amssymb,amsfonts}%
\usepackage{amsthm}%
\usepackage{mathrsfs}%
\usepackage[title]{appendix}%
\usepackage{xcolor}%
\usepackage{textcomp}%
\usepackage{manyfoot}%
\usepackage{booktabs}%
\usepackage{algorithm}%
\usepackage{algorithmicx}%
\usepackage{algpseudocode}%
\usepackage{listings}%
\theoremstyle{thmstyleone}%
\theoremstyle{thmstyletwo}%

\theoremstyle{thmstylethree}%

\begin{document}

\title[Article Title]{Optical Waveguide-based Spider Web Enables Resilient Impact Detection and Localization}


\author[1]{\fnm{Dylan} \sur{Wilson}}

\author*[1]{\fnm{Marco} \sur{Pontin}}\email{marco.pontin@eng.ox.ac.uk}

\author[1]{\fnm{Peter} \sur{Walters}}

\author*[1]{\fnm{Perla} \sur{Maiolino}}\email{perla.maiolino@eng.ox.ac.uk}

\affil[1]{\orgdiv{Department of Engineering Science}, \orgname{University of Oxford}, \orgaddress{\street{Wellington Square}, \city{Oxford}, \postcode{OX1 2JD}, \state{England}, \country{United Kingdom}}}


\abstract{Spiders use their webs as multifunctional tools that enable capturing and localizing prey and more general environmental sensing through vibrations. Inspired by their biological function, we present a spider web-inspired optical waveguide system for resilient impulse detection and localization. The structure consists of six clear thermoplastic polyurethane (TPU) waveguides arranged radially and interconnected by a spiral TPU thread, mimicking orb spider webs. Light transmission losses, induced by vibrations, are measured via coupled LEDs and photo-diodes, allowing real-time detection. We systematically characterize individual waveguides, analyzing key parameters such as tension, impulse position, and break angle to optimize vibrational response. The complete system is validated through controlled experiments, revealing a 5\,ms propagation delay in vibration transfer between adjacent radii, enhancing localization capabilities. We demonstrate a robust impulse detection and localization algorithm leveraging time delay analysis, achieving reliable event identification even in cases of sensor failure. This study highlights the potential of bioinspired optical waveguide structures for adaptive sensing, with applications in soft robotics, structural monitoring, and environmental sensing.}




\maketitle

\section{Introduction}\label{sec1}

For decades, spiders and their remarkable abilities have captivated biologists and engineers. In particular, the complexity and diversity of their webs have spurred research into their mechanics and effectiveness \cite{eberhard1990function}. Spiders have been observed to change the morphology of their webs as they mature \cite{2012_anotaux_AgeingAltersSpider, eberhard1990function}, as well as to adjust the mechanical properties of their silk for different parts of the structure, to achieve specialized functionality and enable functional and structural resilience to damage \cite{2019_mortimer_SpidersVibrationLandscape}. The web acts as a multifunctional tool: it not only allows the spider residing at its center to capture prey and localize it, but it also serves as a way of sensing possible mates and abiotic sources such as wind or nearby sounds \cite{frohlich1982transmission, zhou2022outsourced}. Acting as a large-area sensing device, the web converts physical interactions into vibrations, which are detected by specialized organs in arachnids called slit sensilla \cite{2012_barth_SpiderStrainDetection, barth1998vibrational}. This vibrational behavior is central to the web’s role as a sensing mechanism, contributing to both the spider's survival and its ability to adapt to environmental changes \cite{frohlich1982transmission}.

\begin{figure}
    \centering
    \includegraphics[width=0.9\textwidth]{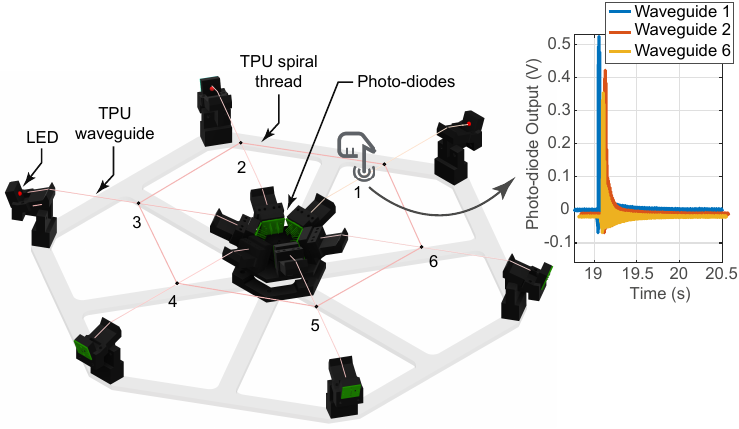}
    \caption{\textbf{Synoptic view of the system}. The optical waveguide spider web consists of six equally spaced clear TPU threads constituting the radii of the web, interconnected by a TPU spiral thread. The web design is inspired by the orb spider's web. LEDs are used to inject light in each waveguide and photo-diodes on the other end are used to measure the light that passes through them. When an impulse is applied to one of the radii, the web vibrates, causing oscillating optical losses in the system, in sync with the mechanical vibrations. These can be measured and analyzed for vibration detection and localization.}
    \label{fig:fig1}
\end{figure}

These features make spider webs the ideal playground for the fields of morphological computation and embodied intelligence, specifically when it comes to understanding how their morphology shapes and filters the sensory output, pre-processing information for the spider, the brain at the core of the system. In this framework, the web takes on an active role in the sensing process, opening endless possibilities for research in the area of morphology optimization \cite{eberhard1990function, frohlich1982transmission}. Previous studies have investigated the possibility of using the webs as media for physical reservoir computing \cite{sadati2018toward} and web-like designs have been employed in the development of metamaterials and nanocomposites \cite{miniaci2016spider, lin2021spider}. In addition, spider webs have served as inspiration for the development of soft sensors, using a range of manufacturing techniques and transduction technologies, for force, tactile and vibration sensing. Resistive technologies in particular have been widely employed. In \cite{zhang20243d}, the authors used Fused Deposition Modelling (FDM) to 3D print strain sensing spider webs using a custom material made by mixing TPU and carbon black powders in a 10\% weight ratio, which was then heated and extruded to create the 3D printing filament. In \cite{huang2018spider} and \cite{liu2018spider}, the authors respectively presented a web-inspired graphene tactile sensor and a wearable and stretchable strain sensor, while in \cite{zhao2021spider} a flexible tactile sensor capable of measuring both pressure and strain is presented. A different approach was adopted by Naderinejad et al. \cite{2023_naderinejad_ExplorationDesignSpiderwebInspired}, who used accelerometers positioned in key areas of 3D printed spider webs to compare the effectiveness of various morphologies for vibration detection and localization. Among the different sensing technologies, methods based on optical waveguides remain largely unexplored, despite their promising potential for vibration detection and localization thanks to their ease of integration and large bandwidth.

In this work, we propose the development of spider webs made out of clear TPU optical waveguides capable of resilient impulse detection and localization (Fig. \ref{fig:fig1}). Each TPU thread is coupled to an LED and a photo-diode. When the TPU waveguide vibrates, light transmission loss occurs and is sensed by the photo-diode, making it possible to pick up vibrations in the system. The addition of a spiral thread enables the coupling of the radii, increasing the sensitivity of the system, localization performance, and resilience of the sensing. In the study, the TPU waveguides are first characterized individually, to capture the effect of key system parameters on their sensory output, followed by the validation of the full system.
\section{Methods}
\begin{figure}
    \centering
    \includegraphics[width=\textwidth]{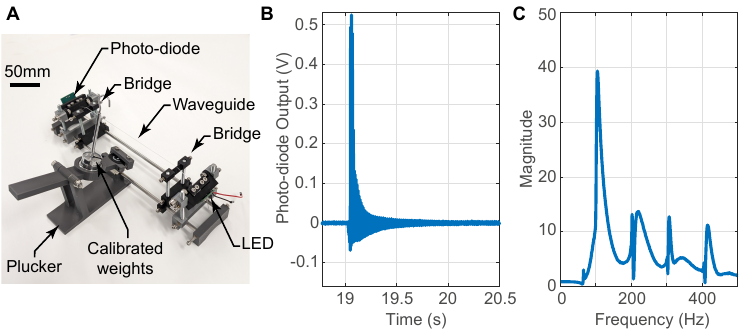}
    \caption{\textbf{Fiber characterization setup}. \textbf{(A)} The physical setup allows for secure clamping of the fiber and adjustment of the break angle at the bridges, while maintaining the vibrating length of the waveguide constant. A plucker device allows for controlled and repeatable impulses to be imparted to the waveguide. \textbf{(B)} and \textbf{(C)} Example acquisition of the measured output from the photo-diode following a pluck and its Fourier transform. For the experiment, a fiber tensioned to 50\,g was used with a break angle of 20$^\circ$ and an impulse imparted using 50\,g calibrated weights.}
    \label{fig:fig2}
\end{figure}
The proposed system is depicted in Fig. \ref{fig:fig1}. The morphology of the device is inspired by that of orb webs. Six equally spaced radii connect the center of the structure to an outer frame. Each of these consists of 0.8\,mm-diameter tensioned clear TPU thread whose ends are coupled to a red LED (621 nm) at the distal end, which shines light into the fiber, and a photo-diode and amplifier (OPT101P Texas Instruments, US) at proximal end, which measures the transmitted light coming out of it. The ends of the waveguides are held in place through 3D printed fixtures, which also align them with the electronics to guarantee good optical coupling. The outputs of the six photo-diodes are acquired by a  USB-6341 DAQ (National Instruments, US), with a sampling frequency of 10\,kHz, necessary to accurately reconstruct the vibrational information expected from our system, whose fibers' natural frequency was measured through preliminary experiments to be around 100\,Hz. The choice of the specific thread material and diameter stemmed from wanting the system to have sustained oscillations and good signal to noise ratio, which make vibration detection easier. In this respect, commercial PMMA fibers are optically superior, but their mechanical properties do not allow for low-frequency vibrations of large amplitude to be achieved. Special elements, similar to bridges in stringed instruments, at both ends of the fiber allow for precise control of the vibrating length and break angle of the fiber, to aid with vibration detection. The length of each vibrating portion of the fibers is 17.5\,cm, in line with common orb spider's web dimensions observed in nature, and its tension was controlled while clamping it, by hanging calibrated weights to it before cutting it to length. All the components of the system aside from the base were 3D printed on a Bambu X1C (Bambu Lab, China) printer in polylactic acid (PLA) (manufacturer). Screws and threaded inserts are used for easy assembly and adjustment of the system and to couple it to the base, laser cut out of 5\,mm acrylic.

When testing individual fibers, a setup replicating one radii of the full system was employed (Fig. \ref{fig:fig2}A). Additionally, a plucking device was devised to allow for controllable impulses to the fiber tested. The hinged horizontal element can be propped up into a repeatable configuration and has a dedicated slot for calibrated weights to fit into. A delrin guitar plectrum is fixed at one end and interacts with the fiber on the downward trajectory. By increasing the weights, the user can impart different angular accelerations to the plucker, allowing for precise and repeatable impulses to the fibers.

\section{Results}\label{sec2}
\subsection{Characterization of individual fibers}
Before assembling the full system, key tests were conducted on individual fibers to test for repeatable behavior and to understand the influence of key parameters on sensor output. Fig. \ref{fig:fig2}B shows the raw output of the photo-diode from one characteristic experiment. The raw output is processed to evaluate the amplitude of the first detected peak, while its Fourier transform is used to analyze the frequency spectrum of the signal, enabling the identification of characteristics such as the mechanical natural frequency of the waveguide.
\begin{figure}
    \centering
    \includegraphics[width=\textwidth]{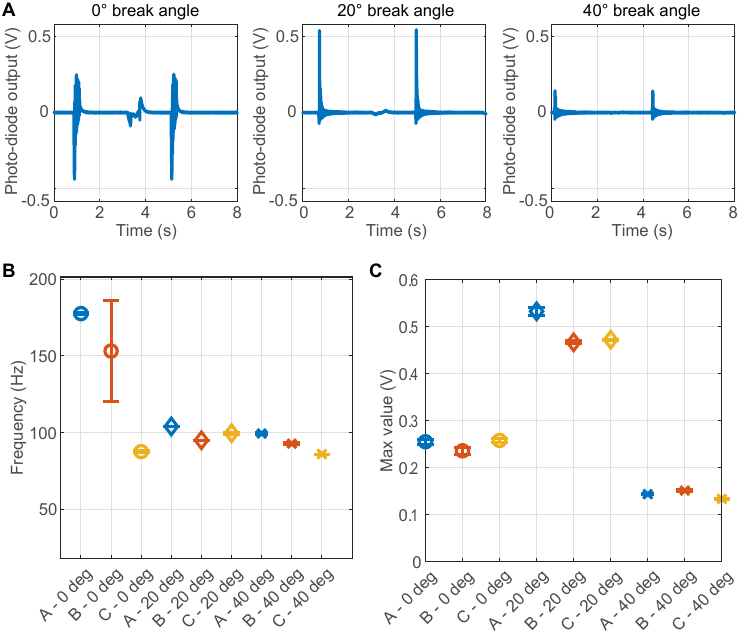}
    \caption{\textbf{Break angle analysis}. \textbf{(A)} Raw data from a representative experiment on one of the waveguides. \textbf{(B)} Shallower break angles cannot reliably isolate the vibration to the portion of the fiber between the two bridges, leading to higher dispersion of data when analyzing the natural frequency of the system. \textbf{(C)} Experimental results reveal that a break angle of 20$^\circ$ grants the best performance from the system, defined as the maximum value of the first peak detected after the waveguide was plucked. In the experiments, three identical waveguides, A, B and C, were used and tested five times, to compute the standard deviations.}
    \label{fig:fig3}
\end{figure}
\subsubsection{Effect of the break angle}
The first experiment analyzed the effect of the break angle on the system. To do this, three identical fibers were positioned in the test rig and the bridges were adjusted to obtain break angles of 0$^\circ$, 20$^\circ$, and 40$^\circ$, while maintaining the vibrating length of the waveguide constant (17.5\,cm). A calibrated weight of 50\,g was applied to the plucker when imparting impulses to the system.

The results of the experiment are displayed in Fig. \ref{fig:fig3}. At a shallow break angle of 0$^\circ$, it was impossible to reliably restrict the vibration to the portion of the waveguide between the two bridges. This resulted in artifacts and additional noise in the raw signal from the photo-diode (Fig.\ref{fig:fig3}A) and large variations in the measured natural frequency for the different fibers (Fig. \ref{fig:fig3}B). The steeper angle of 40$^\circ$ has a positive effect in defining the natural frequency of the waveguide and avoiding signal artifacts, but the peak amplitude is decreased due to the additional optical losses created at the bridge (Fig. \ref{fig:fig3}C). A break angle of 20$^\circ$ seemed to provide a sweet spot that resulted in good natural frequency definition and clean signal, while also allowing for improved sensitivity, with peak amplitudes double those observed with 0$^\circ$ and 20$^\circ$ break angles. For this reason, a break angle of 20$^\circ$ was used in the remainder of this study.

\subsubsection{Effect of the waveguide tension and impulse strength}
\begin{figure}
    \centering
    \includegraphics[width=\textwidth]{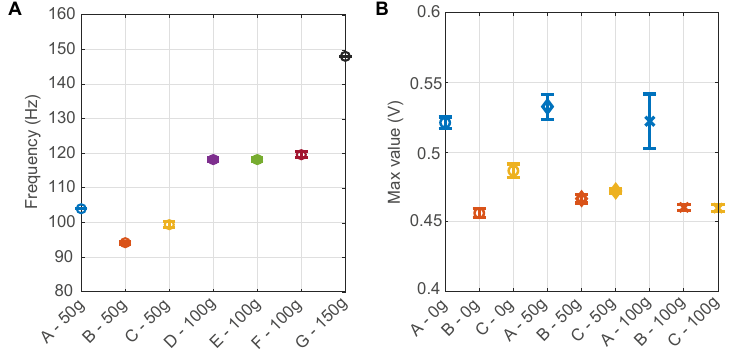}
    \caption{\textbf{Waveguide tension and impulse strength analysis}. \textbf{(A)} The natural frequency of the waveguide can be controlled by changing its tension when locking it in place. In our case, the different tensions were achieved by hanging calibrated weights of different values to one end of the fiber before clamping it and cutting it to length. Three fibers were tested for the lower two tension values and one for the highest. Each was tested five times to get averages and standard deviations in the figure. \textbf{(B)} Impulses of greater and greater magnitude were imparted to the system by adding calibrated weights to the plucker. Three fibers were tested five times for each impulse strength. The results do not highlight any significant relationship between the intensity of the pluck and the amplitude of the first detected peak (one way ANOVA test was performed by pooling together the data from all fibers at each impulse strength values. The test results in a p-value of 0.67).}
    \label{fig:fig4}
\end{figure}
During the study, the length of the waveguide, from LED to photo-diode was kept constant, while the tension on the waveguide was altered with the procedure described in section \ref{sec1}, to study its effect on the natural frequency of the system. Calibrated weights of 50\,g, 100\,g and 150\,g were used to create 6 waveguide samples. As expected, an increase in tension of the waveguide leads to higher measured natural frequencies, with consistent and repeatable behavior, as visible in Fig. \ref{fig:fig4}A.

When considering impulse strength (Fig. \ref{fig:fig4}B), one would expect the amplitude of the first detected peak to strictly relate to the amplitude of the exiting impulse. Tests using the plucker and weights up to 100\,g seem to suggest otherwise, with no discernible trends. This could be due to the way the plucker excites the waveguide: on the downward trajectory, larger weights provide higher angular acceleration, but this might have little effect on the maximum displacement that the waveguide undergoes before slipping off the tip of the plectrum and vibrating. Despite this, we believe our device is still capable of providing a repeatable way of simulating objects falling onto the web at different speeds, in line with the goal of this study.

\subsubsection{Effect of impulse position on a single waveguide}
\begin{figure}
    \centering
    \includegraphics[width=\textwidth]{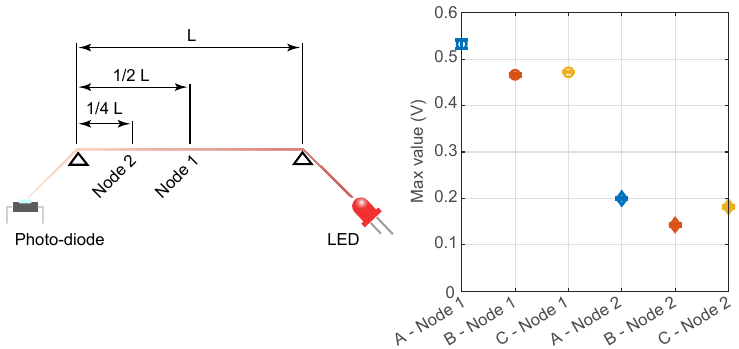}
    \caption{\textbf{Single fiber localization experiment}. Three identical waveguides were plucked five times each in the center and one quarter of the distance away from one of the bridges. The results reveal a statistically significant dependency of the amplitude of the first detected peak of the recorded signal to the position. In particular, plucking with the same intensity closer to the bridge leads to a decrease of the detected peak, which could serve as a way of localizing interactions along the length of the waveguide itself.}
    \label{fig:fig5}
\end{figure}

The last characterization experiment conducted on individual waveguides consists in the possibility of detecting where, along the vibrating portion of the fiber, the impulse was applied. To test for this possibility, three different waveguides, tensioned at 50\,g, were plucked in the center of the vibrating length, Node 1, and a quarter of the way off one of the bridges, Node 2 (left side of Fig. \ref{fig:fig5}). For each node and TPU fiber, five impulses were applied using 50\,g of calibrated weights. As visible in the right side of the figure, plucking the fiber at Node 2 led to consistently smaller amplitudes of the first peak of the raw signal output from the photo-diode. The results therefore suggest the possibility of using the system not only to locate which fiber was plucked in a web, but also where along the fiber the interaction occurred.
\subsection{Validation of the optical waveguide web}
\begin{figure}
    \centering
    \includegraphics[width=0.9\textwidth]{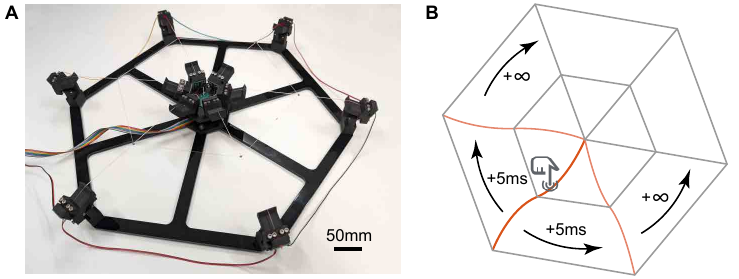}
    \caption{\textbf{The complete system}. \textbf{(A)} Photograph of the real system. \textbf{(B)} When plucking a fiber, the spiral thread transfers the vibration across to the adjacent radii. Experimental validation showed that there is a 5\,ms delay on average between the plucked fiber signal first peak and those of the waveguides directly adjacent to it. Moving further along, the low signal to noise ratio makes it difficult to precisely detect peaks associated with the propagation of vibrations from the plucked fiber.}
    \label{fig:fig6}
\end{figure}

The full system was built using only pre-characterized fibers resulting in one 50\,g-tension waveguide, four 100\,g-tension ones and one 150\,g-tension waveguide for radii. A TPU spiral thread, made out of the same thread as the radii, was added to couple the radii together. The spiral thread was locked in place, as to split each radius in half, using small elastic hoops. These were selected to avoid scratching the delicate TPU threads, which would compromise light transmission. With the setup assembled, a dataset was collected by plucking by hand each fiber where it met the spiral thread five times. The data was then processed to detect the first peak on each fiber. The analysis showed that, following an impulse at time $t_0$ on one fiber, peaks on the two nearest neighbors appear on average $t_0+$5\,ms ($SD=$0.8\,ms). This result is summarized in Fig. \ref{fig:fig6}B, where the $\infty$ symbol is due to the fact that the attenuation of the vibration did not allow for reliable detection of peaks on fibers further away.

\begin{figure}
    \centering
    \includegraphics[width=0.9\textwidth]{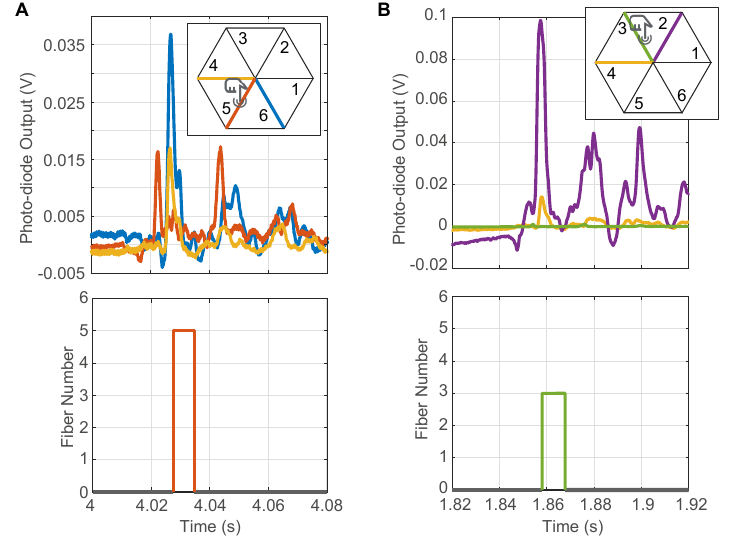}
    \caption{\textbf{Impulse detection and localization}. \textbf{(A)} Example of impulse localization when all fibers involved in the localization process are working. Only the plucked fiber and its nearest neighbors are shown for clarity. The localization is straightforward as the peak on the plucked fiber appears first, followed by peaks on the adjacent fibers around 5\,ms later. \textbf{(B)} Example of localization experiment that relies on the built-in resilience of the system. When an impulse happen on a non-functioning fiber, the localization can still happen successfully, as two simultaneous peaks are detected on two non-adjacent fibers. The algorithm therefore successfully infers that the fiber in between the two was plucked (fiber 3 in this case, as highlighted in the bottom chart).}
    \label{fig:fig7}
\end{figure}

The existence of the aforementioned time delay in the appearance of the peaks in the photo-diodes output signals, due to the mechanics of the system, allows for the development of a naive yet robust impulse detection and localization algorithm. A threshold is set, based on the training data, that, when surpassed by any of the waveguide outputs, triggers the impulse detection. A further 100 samples are collected (corresponding to 10\,ms of data) for all fibers and then the \textit{findpeaks} function in MATLAB is used to detect the first peak in each signal. The time $t_0$ of the first detected peak overall is then subtracted from every other to compute six time deltas, one of which is always zero by definition. If no peaks are detected on a fiber, a value of $NaN$ is assigned to it. If all fibers work properly, the localization problem is trivial: the fiber being plucked corresponds to the time delta of 0\,ms and will have 5\,ms deltas left and right of it. This case is demonstrated in Fig. \ref{fig:fig7}A, together with the corresponding successful localization.

Thanks to the presence of the spiral thread, though, more complex cases can also be resolved. In particular, the fact that the vibrations from one radial thread are transmitted to the adjacent ones, allows for the system to be resilient in case one fiber stops working. Assuming the plucked fiber is not responding, two almost simultaneous peaks ($<$3.5\,ms apart) are detected on non-adjacent fibers. When this is the case, the algorithm concludes that the waveguide in between those two has been plucked. We tested this case by artificially scaling the output of the fiber being plucked by a factor of 0.1. As visible in Fig. \ref{fig:fig7}B, this leads to the disappearance of the peak on fiber number 3. Still, the system is capable of detecting the peaks on fibers 2 and 4, which happen almost simultaneously, and from there conclude that fiber 3 had been plucked.

\section{Discussion and Conclusions}\label{sec3}
The remarkable properties of spider webs, and the multi-functionality they provide to the spider at their center have inspired researchers for decades. The variety of web morphologies further suggests the web may actively contribute in the pre-processing of information coming from physical interactions, before this reaches the vibration-sensitive organs of these animals.

In this work, we explored the possibility of creating spider web-inspired sensors, based on soft optical waveguides, for impact detection and localization. TPU threads were chosen for their elastic properties and good light transmission capabilities. Characterization of individual fibers allowed us to assess the effect of key parameters on the sensory output. In particular, the analysis allowed us to define a 20$^\circ$-break angle as the best one for the fiber being tested. In addition, altering the tension of the waveguide allows for controlling its natural frequency and therefore of the oscillatory frequency of the photo-diode output. This allows the system designer to place the impulse response of each waveguide within a desired range, potentially allowing for more accurate detection and localization of vibrations. Further analysis showed that it can be possible to reconstruct where along a fiber the impulse is provided, by looking at the amplitude of the first peak in the photo-diode output. Moving to the full system, the spiral thread enables the coupling of the six radial threads, further improving the performance of the system. Thanks to the propagation delay in vibration transfer within the system, we demonstrated the possibility of using a naive localization approach which proved robust even in case one of the waveguides stops working, making the sensing resilient to single-point failures.

Future work will expand on the signal processing by applying machine learning techniques to improve detection and localization and expand it to multiple simultaneous impulses and multiple failures. Larger training datasets might also lead to the localization being capable of detecting impulses applied to the spiral thread, as well as tell where, along one fiber, the impulse occurred. In the current system, one of the challenges consists in manufacturing fibers that have similar base responses. This is due to the fact that the TPU thread employed in the study was not manufactured for use as an optical waveguide. As such, no information is given by the manufacturer on optical or mechanical properties. As commercial soft, elastic fiber optics become more readily available, this problem should be easily overcome. An alternative method, which we intend to investigate in the future, is the possibility of direct printing of the desired web morphology out of clear TPU, which would allow for faster and more repeatable manufacturing as well as a higher degree of customization of the sensor.

\section{Acknowledgments}
This work was supported by Engineering and Physical Sciences Research
Council (EPSRC) Grant EP/V000748/1.

\bibliography{sn-bibliography}% common bib file

\end{document}